\documentclass[%
 reprint,
 amsmath,amssymb,
 aps,
]{revtex4-2}

\usepackage{graphicx}% Include figure files
\usepackage{dcolumn}% Align table columns on decimal point
\usepackage{bm}% bold math
\usepackage{float}
\newcommand{\Romannum}[1]{\MakeUppercase{\romannumeral #1}}
\begin{document}

\preprint{APS/123-QED}

\title{Quantitative Determination of Quantum Fluctuations in Clean Magnets \\ \Romannum{1}: Neutron Spin Echo}% Force line breaks with \\
%\thanks{uncompleted draft}%

\author{Fanjun Xu}
 \email{fjxuphys@gmail.com}
 \affiliation{Helmholtz-Zentrum Berlin für Materialien und Energie GmbH, Hahn-Meitner Platz 1, D-14109 Berlin, Germany}%
 %\affiliation{Institut für Festkörperphysik, Technische Universität Berlin, Hardenbergstraße 36, D-10623 Berlin, Germany}

\date{\today}% It is always \today, today,

\begin{abstract}
%Starting from the fundamental prisnciples of magnetic neutron scattering, we show that the level of quantum fluctuations in quantum and frustrated magnets can be quantitatively determined from the ratio of inelastic to total magnetic spectral weight. Exploiting the magnetic total-moment sum rule, we establish a direct relationship between the long-time limit of the intermediate scattering function and the ratio of ordered to total magnetic moments,

Starting from the magnetic total-moment sum rule of neutron scattering, we derive an explicit connection between ordered-moment reduction and the long-time limit of the intermediate scattering function. We show that this time-domain formulation establishes a direct and experimentally accessible measure of quantum fluctuation strength through neutron spin-echo spectroscopy, with

\[P(t\rightarrow\infty)=\frac{I(Q,t\rightarrow\infty)}{I(Q,t=0)}=\frac{\langle \mu  \rangle^{2}}{\langle \mu^{2}  \rangle}\]

This identity links the long-time polarization to the ratio between ordered and total magnetic moments. Linear spin-wave calculations for square and triangular Heisenberg antiferromagnets demonstrate that both quantum effect and geometric frustration suppress the plateau value in quantitative agreement with moment reduction. The resulting framework establishes a direct and model-independent measure of the level of quantum fluctuations in bulk quantum magnets, particularly for polycrystalline samples.

\end{abstract}

\maketitle

%\tableofcontents

%\section{\label{sec:level1}INTRODUCTION}
\section{INTRODUCTION}

Quantum magnets differ fundamentally from their classical counterparts due to intrinsic zero-point fluctuations arising from the non-commuting nature of spin operators. These quantum fluctuations are enhanced in systems with small spin magnitude, reduced dimensionality, or competing exchange interactions, such as geometrically frustrated lattices. A central problem in experimental quantum magnetism is therefore to quantify the extent to which magnetic moments are reduced by fluctuations, and to determine how spectral weight is redistributed between static and dynamical correlations \cite{lacroix2011introduction}.

Compared with isolated atoms or few-spin systems, quantifying the degree of “quantumness” in bulk condensed-matter systems is substantially more challenging. In real materials, the many-body ground state is embedded in an extended statistical ensemble, and experimentally measurable observables are collective in nature. Consequently, signatures of quantum fluctuations are typically inferred indirectly or extracted through model-dependent analysis. Establishing a direct, model-independent experimental quantifier of quantum fluctuation strength in bulk magnets is therefore highly desirable.

A direct manifestation of quantum fluctuations in solids is the reduction of the ordered magnetic moment together with the redistribution of spectral weight over energy transfer. In this work, we show that neutron spin echo (NSE) spectroscopy \cite{mezei1972neutron} provides a natural time-domain framework to quantify this redistribution in bulk quantum magnets, particularly in polycrystalline samples. By focusing on the ordered moment as an experimentally accessible quantity in neutron scattering, we show that the long-time polarization measured by NSE directly quantifies the degree of ordered-moment reduction.

In the following, we develop this connection starting from the magnetic total-moment sum rule and illustrate its implications using linear spin-wave calculations for prototypical Heisenberg antiferromagnets. Our discussion is restricted to clean local-moment systems at sufficiently low temperatures, where the suppression of the ordered moment is dominated by intrinsic zero-point quantum fluctuations.

\section{Total Moment Sum Rule Revisited}

For localized magnetic moments in Mott insulators, the magnetic neutron scattering intensity follows the total moment sum rule \cite{lorenzana2005sum,boothroyd2020principles}:

\begin{equation}
\int_{\text{BZ}} d^3Q \int d(\hbar\omega)\,
S(\mathbf{Q},\omega)
=
\frac{2}{3}\,\langle \mu^2 \rangle .
\label{eq:sumrule_total}
\end{equation}

The integration is performed over one Brillouin zone in reciprocal space and over all energy transfers. 
The prefactor $2/3$ arises from the polarization factor of magnetic neutron scattering, reflecting the powder-averaged projection of spin components perpendicular to $\mathbf{Q}$, which contributes a factor of $2/3$.

The total magnetic spectral weight can be separated into elastic and inelastic contributions:

\begin{equation}
S(\mathbf{Q},\omega)
=
S_{\text{el}}(\mathbf{Q})\,\delta(\omega)
+
S_{\text{inel}}(\mathbf{Q},\omega).
\end{equation}

Integrating only the inelastic part yields

\begin{equation}
\int_{\text{BZ}} d^3Q \int d(\hbar\omega)\,
S_{\text{inel}}(\mathbf{Q},\omega)
=
\frac{2}{3}
\left(
\langle \mu^2 \rangle - \langle \mu \rangle^2
\right),
\label{eq:sumrule_inel}
\end{equation}

which corresponds to the square of the fluctuating magnetic moment.

Therefore, the fraction of inelastic magnetic spectral weight directly measures the relative strength of quantum fluctuations:

\begin{equation}
\frac{\sigma_{\text{inel}}}{\sigma_{\text{total}}}
=
1 - \frac{\langle \mu \rangle^2}{\langle \mu^2 \rangle}.
\label{eq:fluctuation_ratio}
\end{equation}

This provides a model-independent definition of the degree of moment reduction due to quantum fluctuations.

\section{Experimental Protocol with Neutron}

In practice, separating elastic and inelastic magnetic spectral weight requires careful experimental considerations. Ideally all measurement mentioned below should be done with polarization analysis to disentangle the pure magnetic component from non-magnetic scattering.

%In practice, ideally all measurement mentioned below should be done with polarization analysis to disentangle the pure magnetic component from non-magnetic scattering.

\subsection{Conventional Diffraction and Spectroscopy}

Neutron diffraction measures the elastic magnetic structure factor 
$S(\mathbf{Q},\omega=0)$, this yields information on the ordered moment $\langle \mu \rangle$. In practice, conventional neutron diffractometers do not employ energy analysis; therefore the measured intensity corresponds to $S(\mathbf{Q})$ integrated over all energy transfer.

Energy-resolved neutron spectroscopy, such as time-of-flight (TOF) or triple-axis spectroscopy (TAS), measures the dynamical structure factor 
$S(\mathbf{Q},\omega)$ over a finite energy window. By integrating the magnetic intensity over energy within this window, the total spectral weight can be estimated. The elastic and total magnetic scattering can be distinguished by energy analysis.

However, in practice the separation between elastic and inelastic scattering is fundamentally limited by the instrumental energy resolution $\Delta E$. A finite energy resolution corresponds to a finite accessible time window via the uncertainty relation:

\begin{equation}
t_{\max} \sim \frac{\hbar}{\Delta E}.
\end{equation}

For state-of-the-art neutron spectrometers with energy resolution in the range 
$\Delta E \sim 0.01$--$0.05$~meV,
the corresponding time window is approximately

\begin{equation}
t_{\max}
\sim
\frac{0.658~\text{meV·ps}}{\Delta E}
\approx
10^2 \text{--} 10^1~\text{ps}.
\end{equation}

While Neutron backscattering instruments, with $\Delta E \sim (3\text{--}7.5)\times10^{-4}~\text{meV}$
extend this time window to the nanosecond regime, up to several nanoseconds.

%(or backscattering to around 7 ns with energy resolution 0.0003–0.00075 meV)
Thus, extremely slow fluctuations may appear as elastic (resolution-limited) within the experimental time window, even though they are dynamically fluctuating on longer timescales. The true long-time dynamical behavior beyond the experimental window remains unresolved. This intrinsic limitation complicates a direct experimental evaluation of the fluctuation ratio defined in Eq.~(\ref{eq:fluctuation_ratio}), as the apparent elastic fraction depends explicitly on the instrument resolution.

%This limitation complicates the direct experimental evaluation of the fluctuation ratio in Eq.~(\ref{eq:fluctuation_ratio}).

\subsection{Neutron Spin Echo: at long time limit}

%\section{Intermediate Scattering Function and Long-Time Limit}

The intermediate scattering function is the real part of the Fourier transform of the dynamical structure factor. In neutron spin echo (NSE) experiments, the measured neutron polarization directly approximates the normalized intermediate scattering function \cite{mezei1972neutron,mezei2002neutron}:

\begin{equation}
P(t)
=
\frac{\int S(Q,\omega)\cos(\omega t)\, d\omega}
{\int S(Q,\omega)\, d\omega}.
\label{eq:Pt_def}
\end{equation}

Separating elastic and inelastic contributions,

\begin{equation}
S(Q,\omega)
=
S_{\mathrm{el}}(Q)\,\delta(\omega)
+
S_{\mathrm{inel}}(Q,\omega),
\end{equation}

the polarization can be written as

\begin{equation}
P(t)
=
P_{\mathrm{el}}(t)
+
P_{\mathrm{inel}}(t),
\end{equation}

with

\begin{equation}
P_{\mathrm{el}}(t)
=
\frac{S_{\mathrm{el}}(Q)}
{\int S(Q,\omega)\, d\omega},
\end{equation}

which is time independent, and

\begin{equation}
P_{\mathrm{inel}}(t)
=
\frac{\int S_{\mathrm{inel}}(Q,\omega)\cos(\omega t)\, d\omega}
{\int S(Q,\omega)\, d\omega}.
\end{equation}

For systems in which the inelastic spectral weight is continuous in energy, the Riemann–Lebesgue lemma impliestime

\begin{equation}
\lim_{t\rightarrow\infty} P_{\mathrm{inel}}(t) = 0.
\end{equation}

Therefore, the long-time limit of the measured polarization is governed solely by the elastic spectral weight:

\begin{equation}
\lim_{t\rightarrow\infty} P(t)
=
\frac{S_{\mathrm{el}}(Q)}
{\int S(Q,\omega)\, d\omega}
=
\frac{\sigma_{\mathrm{elastic}}}
{\sigma_{\mathrm{total}}}.
\label{eq:plateau}
\end{equation}

This result applies to systems in which the magnetic scattering is concentrated at $\omega = 0$ (for example, resolution-limited Bragg peaks), while the inelastic spectrum is continuous in energy. In such cases, the inelastic contribution to $P(t)$ typically decays on a time scale set by the inverse bandwidth of the magnetic excitations, often within the picosecond range. At longer times (e.g., $t \gtrsim 10^2$~ps for conventional spectrometers), the polarization is therefore dominated by the elastic contribution, giving rise to a plateau. Extremely slow fluctuations or quasi-static motion may broaden the nominally elastic line. In this situation, the plateau can exhibit a gradual decay at long times, reflecting the finite lifetime associated with these slow dynamics. 

Therefore, within the experimentally accessible time window, the plateau value directly reflects the static (elastic) magnetic spectral weight. Connecting to the total-moment sum rule, the plateau which shows up at long-time can be expressed as

\begin{equation}
\lim_{t\rightarrow\infty} P(t)
=
\frac{\langle \mu \rangle^2}
{\langle \mu^2 \rangle}.
\label{eq:moment_ratio_time}
\end{equation}
which is equivalent to Eq.~(\ref{eq:fluctuation_ratio}) derived in the energy domain.
Thus, the final polarization measured by neutron spin echo provides a direct and model-independent determination of the ratio between ordered and total magnetic moments. Modern cold-neutron wide-angle NSE instruments such as SPAN \cite{pappas2000novel} and WASP \cite{fouquet2007wide} provide direct experimental access to this long-time limit over broader Q space that can cover the 1st Brillouin zone.

Importantly, this approach determines the redistribution of magnetic spectral weight between static and dynamical channels directly from measured neutron polarization. It is model-independent and does not require inversion procedures, fitting to microscopic Hamiltonian, or reconstruction of dynamical susceptibilities, and is therefore broadly applicable to polycrystalline quantum magnets.

\section{Discussion}
To illustrate the proposed protocol quantitatively, we evaluate the long-time polarization within linear spin-wave theory (LSWT) for prototypical Heisenberg antiferromagnets described by

%To illustrate the proposed protocol quantitatively, we evaluate the long-time polarization within linear spin wave theory (LSWT) for two prototypical Heisenberg antiferromagnet based on spin only Hamiltonian described by

\begin{equation}
H = J_1 \sum_{\langle ij\rangle} \mathbf{S}_i \cdot \mathbf{S}_j .
\label{eq:HeisenbergHamiltonian}
\end{equation}

\begin{figure}
    \centering
    \includegraphics[width=1\linewidth]{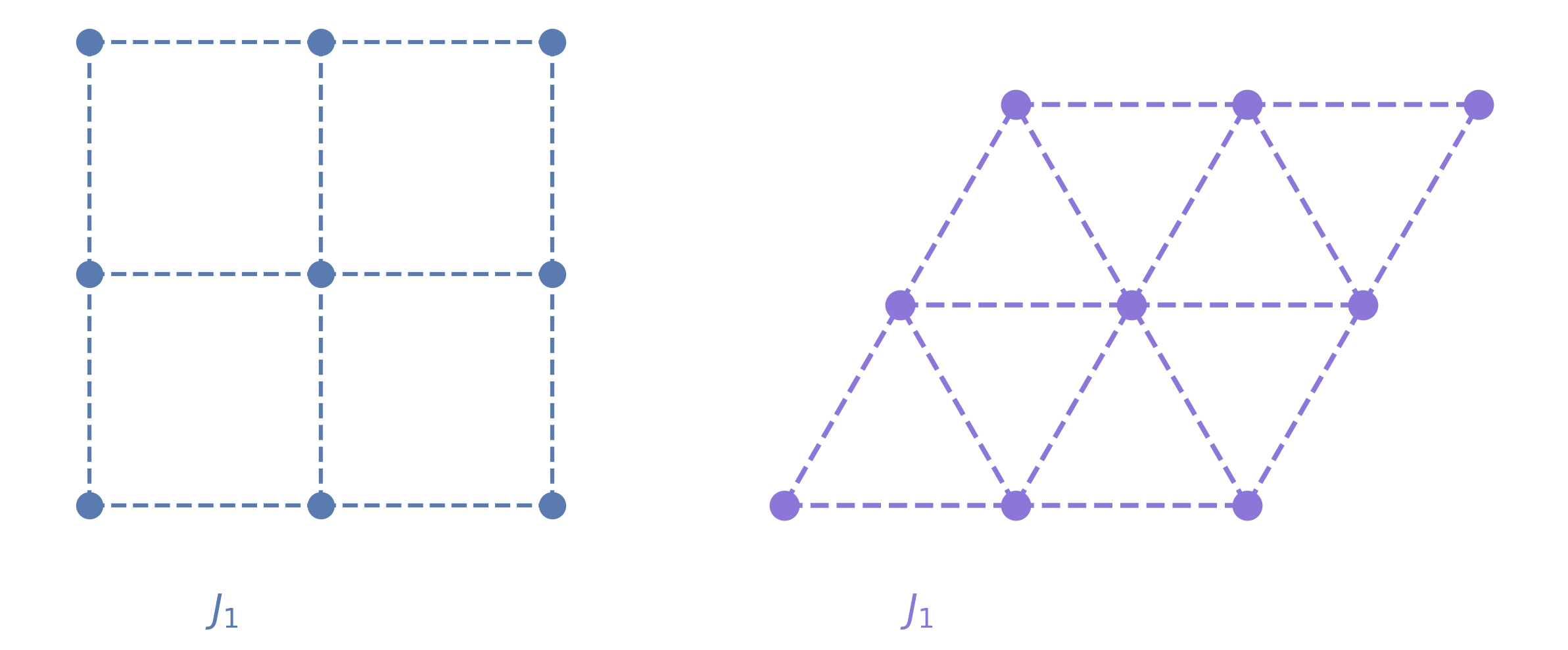}
    \caption{J$_{1}$ only Square lattice and Triangular lattice.}
    \label{fig:SqTriLattice}
\end{figure}

While LSWT does not capture higher-order quantum corrections in strongly frustrated $S=1/2$ systems, it provides a controlled illustration of the relationship between ordered-moment reduction and long-time polarization.

\begin{figure}[h]
    \centering
    \includegraphics[width=1\linewidth]{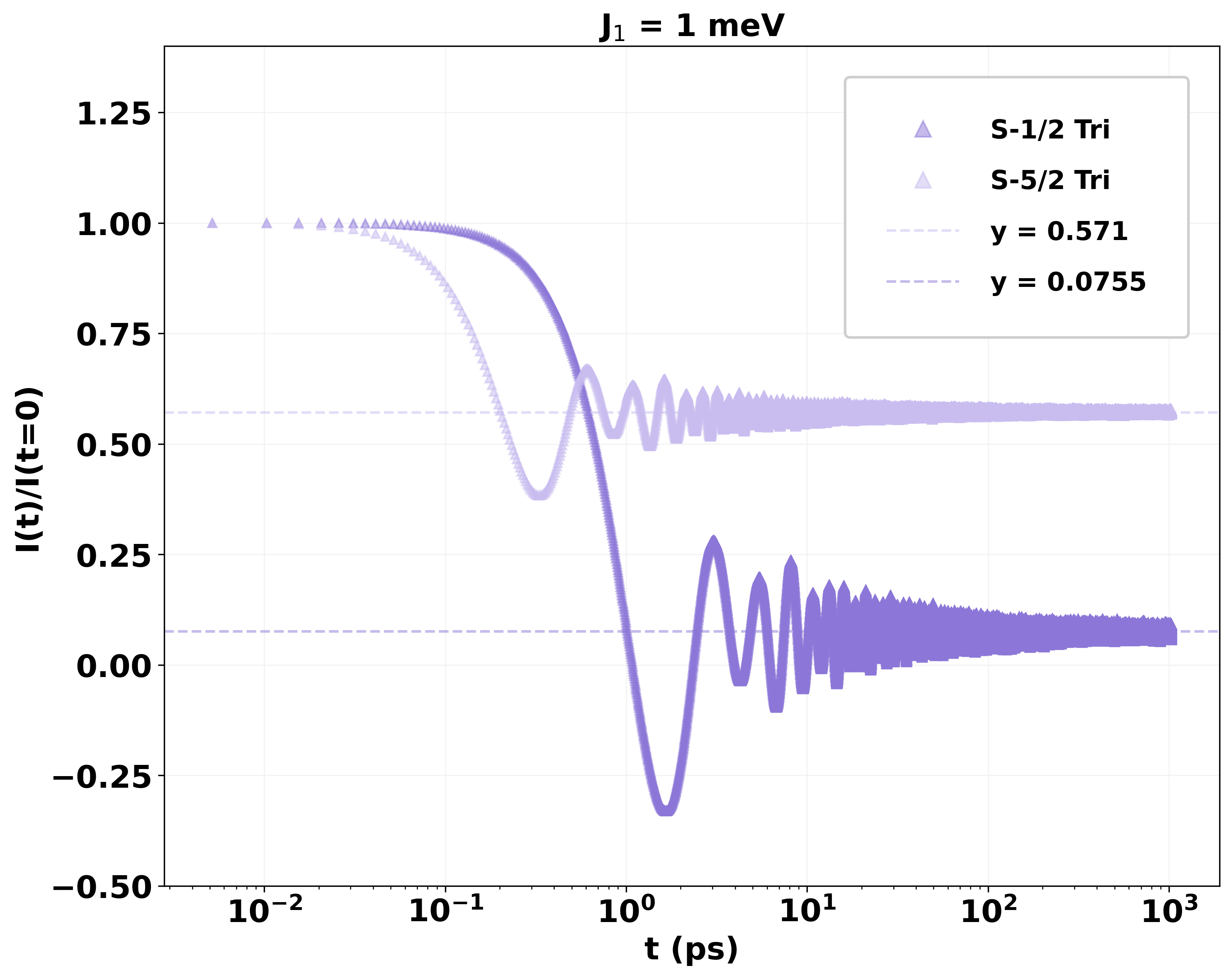}
    \caption{LSWT of triangular lattice with S-5/2 (light purple) and S-1/2 (dark purple).}
    \label{fig:LSWT_Triquantumeffect}
\end{figure}

\begin{figure}[h]
    \centering
    \includegraphics[width=1\linewidth]{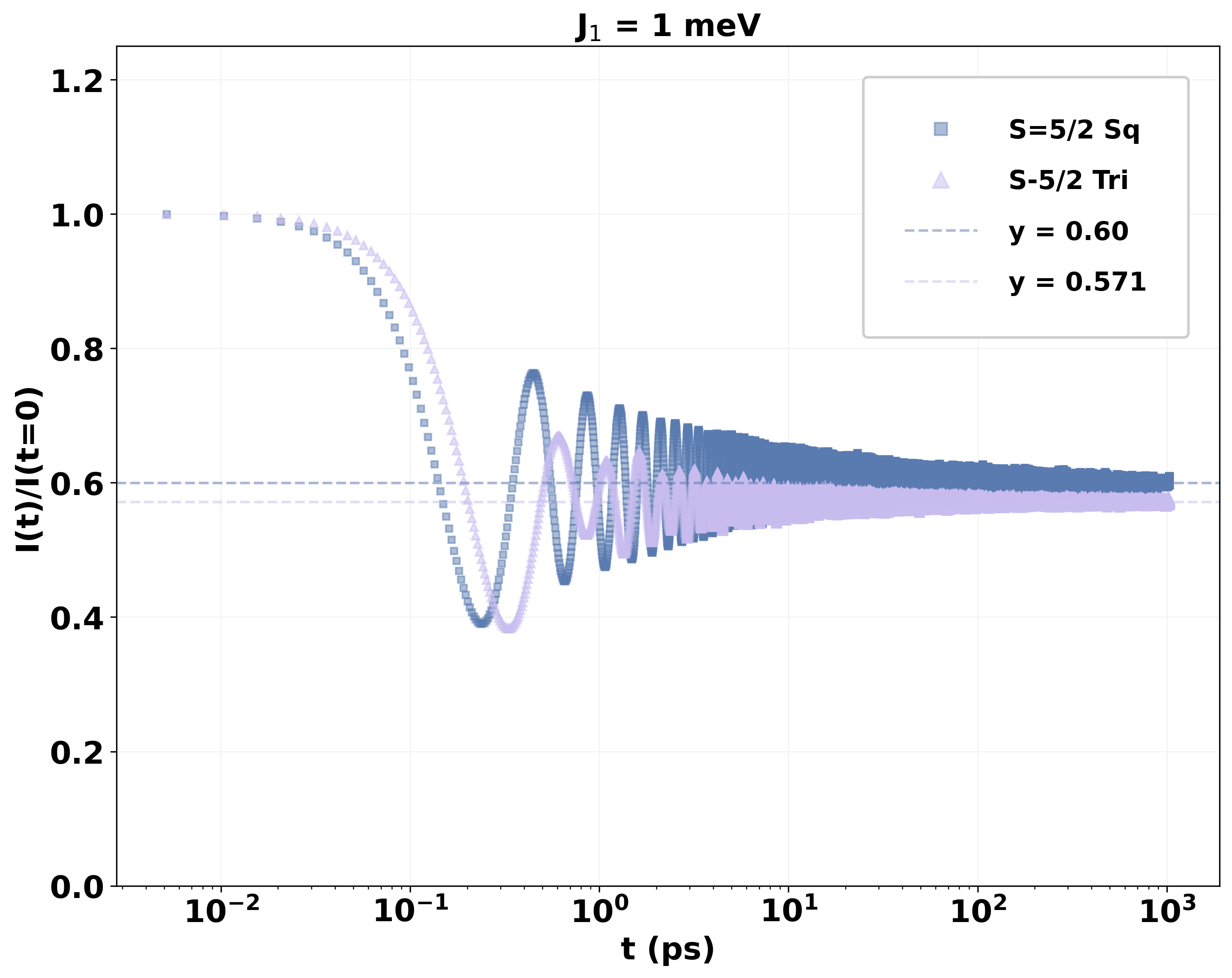}
    \caption{LSWT of Square lattice (blue) and triangular lattice (purple) with S-5/2.}
    \label{fig:LSWT_SquareTri}
\end{figure}

We consider two prototypical lattices: the unfrustrated square lattice and the geometrically frustrated triangular lattice, as shown in FIG.~\ref{fig:SqTriLattice}. For each lattice, we examine both a classical spin value $S=5/2$ and a quantum spin value $S=1/2$. These two lattices provide a minimal comparison between the spin number and geometric frustration. The calculated intermediate scattering functions, illustrating the influence of spin magnitude (quantum versus classical limit) and lattice geometry (unfrustrated versus frustrated) are shown in Figs.~\ref{fig:LSWT_Triquantumeffect} and Figs.~\ref{fig:LSWT_SquareTri}, respectively.

The dynamical structure factor was computed within LSWT using the SpinW package~\cite{toth2015linear}. The magnetic structure was defined with ordering vector $\mathbf{k}=(\tfrac{1}{2},\tfrac{1}{2},0)$ for the square lattice and the corresponding $120^\circ$ structure for the triangular lattice. Powder-averaged spectra were obtained by Monte Carlo sampling of reciprocal-space orientations and evaluated on a dense energy grid. The resulting $S(Q,\omega)$ was integrated over $Q$ to approximate first Brillouin-zone averaging. The time-dependent intermediate scattering function was then obtained by discrete cosine transformation of the energy-resolved spectrum.

The calculated ordered moments $\langle S^z \rangle$ and the corresponding static spectral-weight fractions $\langle S^z \rangle^2 / S(S+1)$ are summarized in Table~\ref{tab:LSWT_results}. The deviation of $\langle S^z \rangle$ from its classical value $S$ provides a quantitative measure of quantum fluctuation strength.

\begin{table}[t]
\caption{Linear spin-wave theory results for the ordered moment and the corresponding plateau ratio 
$\langle S^z \rangle^2 / S(S+1)$ for square (Sq) and triangular (Tri) lattices.}
\label{tab:LSWT_results}
\begin{ruledtabular}
\begin{tabular}{cccc}
Lattice & $\langle S^z \rangle$ & $S$ & $\langle S^z \rangle^2 / S(S+1)$ \\
\hline
$S=\tfrac{1}{2}$ Sq & 0.304 & 0.5 & 0.123 \\
$S=\tfrac{1}{2}$ Tri & 0.238 & 0.5 & 0.0755 \\
$S=\tfrac{5}{2}$ Sq & 2.29 & 2.5 & 0.6 \\
$S=\tfrac{5}{2}$ Tri & 2.23 & 2.5 & 0.572 \\
\end{tabular}
\end{ruledtabular}
\end{table}

The corresponding intermediate scattering functions are shown in Figs.~\ref{fig:LSWT_Triquantumeffect} and Figs.~\ref{fig:LSWT_SquareTri}. In all cases, $P(t)$ exhibits an initial decay associated with inelastic spin-wave excitations, followed by a plateau at long times. The plateau value agrees quantitatively with the LSWT prediction in Table~\ref{tab:LSWT_results}, confirming the direct correspondence between long-time polarization and ordered-moment reduction.

%In all cases, $P(t)$ exhibits a rapid decay associated with inelastic spin-wave excitations, followed by a plateau at long times. The plateau value agrees with the LSWT prediction \ref{tab:LSWT_results}, confirming the direct correspondence between the long-time polarization and the ordered-moment reduction.

%\begin{equation}
%P(t\rightarrow\infty)
%=
%\frac{\langle \mu \rangle^2}{\langle \mu^2 \rangle},
%\end{equation}

Two systematic trends emerge clearly:

\begin{enumerate}
\item \textbf{Quantum spin reduction:} For a fixed lattice geometry, decreasing the spin magnitude from $S=5/2$ to $S=1/2$ lowers the plateau value, reflecting enhanced quantum fluctuations.

\item \textbf{Geometric frustration:} For a fixed spin magnitude, the triangular lattice exhibits a smaller plateau than the square lattice, indicating additional suppression of ordered moment due to frustration.
\end{enumerate}

In the classical limit ($S \gg 1$), the ordered moment approaches its full value and the plateau tends toward unity. In contrast, in a paramagnetic state without elastic magnetic weight, $P(t)$ decays to zero. Systems with partial moment reduction display intermediate plateau values, providing a quantitative measure of fluctuation strength.

Importantly, the presence of a plateau within the experimental time window does not imply the absence of dynamics. Rather, the plateau value reflects the degree of ordered moment reduction, while dynamical fluctuations are encoded in the initial decay of $P(t)$.

\subsection*{Limitations and Practical Considerations}

The protocol applies primarily to clean local-moment systems, such as insulating magnets that obey the magnetic total-moment sum rule. It is particularly suitable for polycrystalline samples, where wide-angle coverage approximates Brillouin-zone (BZ) integration.

Caution is required for itinerant or metallic systems, where magnetic moments are not strictly localized, and for cases with strong quasi-elastic scattering that complicates the separation of static and slow dynamical contributions. The extracted plateau value also depends on the integrated $Q$ range; limited Brillouin zone sampling introduces systematic deviations.

Strict equality to the sum rule requires integration of the intrinsic spin correlation function over the full BZ with uniform $d^3Q$ weighting. In practice, the measured intensity must be corrected for the magnetic form factor, polarization factor prior to integration.

Finally, ordered-moment suppression may arise not only from zero-point quantum fluctuations but also from thermal effects or structural disorder. The observed plateau reduction therefore reflects the combined influence of intrinsic and disorder-induced suppression of static magnetic order.
%In real materials, ordered-moment suppression may arise not only from intrinsic zero-point quantum fluctuations but also from thermal fluctuations: e.g. structural disorder. In such cases, the observed plateau reduction reflects the combined effect of intrinsic quantum fluctuations and any additional classical or disorder-induced suppression of static magnetic order.

\section{Conclusion}

%\section{Conclusion}

We have proposed a direct and model-independent method to quantify the level of quantum fluctuations in local-moment magnets based on the magnetic total-moment sum rule. By separating elastic and inelastic spectral weight and reformulating the problem in the real time domain, we show that the long-time limit of the neutron spin echo intermediate scattering function satisfies

\[
P(t\rightarrow\infty)
=
\frac{\langle \mu \rangle^{2}}{\langle \mu^{2} \rangle},
\]

providing a direct experimental measure of the ratio between ordered and total magnetic moments.

Numerical calculations within linear spin wave theory for square and triangular Heisenberg antiferromagnets demonstrate that both quantum spin reduction and geometric frustration lower the plateau value of the intermediate scattering function, in quantitative agreement with the predicted moment reduction. 

Within the experimentally accessible time window, the plateau of the neutron spin echo signal directly reflects the static magnetic spectral weight. The method is particularly suitable for local-moment systems and can be implemented using wide-angle neutron spin echo spectroscopy with appropriate integration $Q$.

%The framework developed here establishes an operational definition of quantum fluctuation strength in bulk quantum magnets and provides a systematic basis for interpreting neutron spin-echo measurements in frustrated magnetic systems.

\section*{Acknowledgments}

We thank Bella Lake, Margarita Russina, Peter Fouquet, and Björn Fåk for general discussions on neutron spectroscopy. We acknowledge financial support from Helmholtz-Zentrum Berlin. We further acknowledge financial support from the Röntgen-Ångström Cluster (RÅC) for participation in the RÅC International Summer School and from the MaMaSELF Erasmus Mundus program. This manuscript benefited from the use of ChatGPT for language polishing and assistance with code debugging; we are responsible for scientific content, analysis, and conclusions. Simulation codes are available upon reasonable request.

\section{Supplementary Material}
\label{sec:supplement}

%\subsection{Numerical Method}

\subsection{Effect of Elastic Line Broadening}

In practice, the elastic contribution is not strictly a delta function in energy but can acquire a finite intrinsic width due to slow dynamics or instrumental resolution effects. As a consequence, the nominally elastic component produces a gradual decay of the long-time plateau in the intermediate scattering function.

Assuming a Gaussian energy broadening of width $\Delta E$, the elastic contribution in the time domain takes the form
\[
P_{\mathrm{el}}(t)
=
\frac{\sigma_{\mathrm{el}}}{\sigma_{\mathrm{total}}}
\exp\left(-\frac{1}{2}(\sigma t)^2\right).
\]
where $\sigma = \Delta E / \hbar$.

\begin{figure}[H]
    \centering
    \includegraphics[width=1\linewidth]{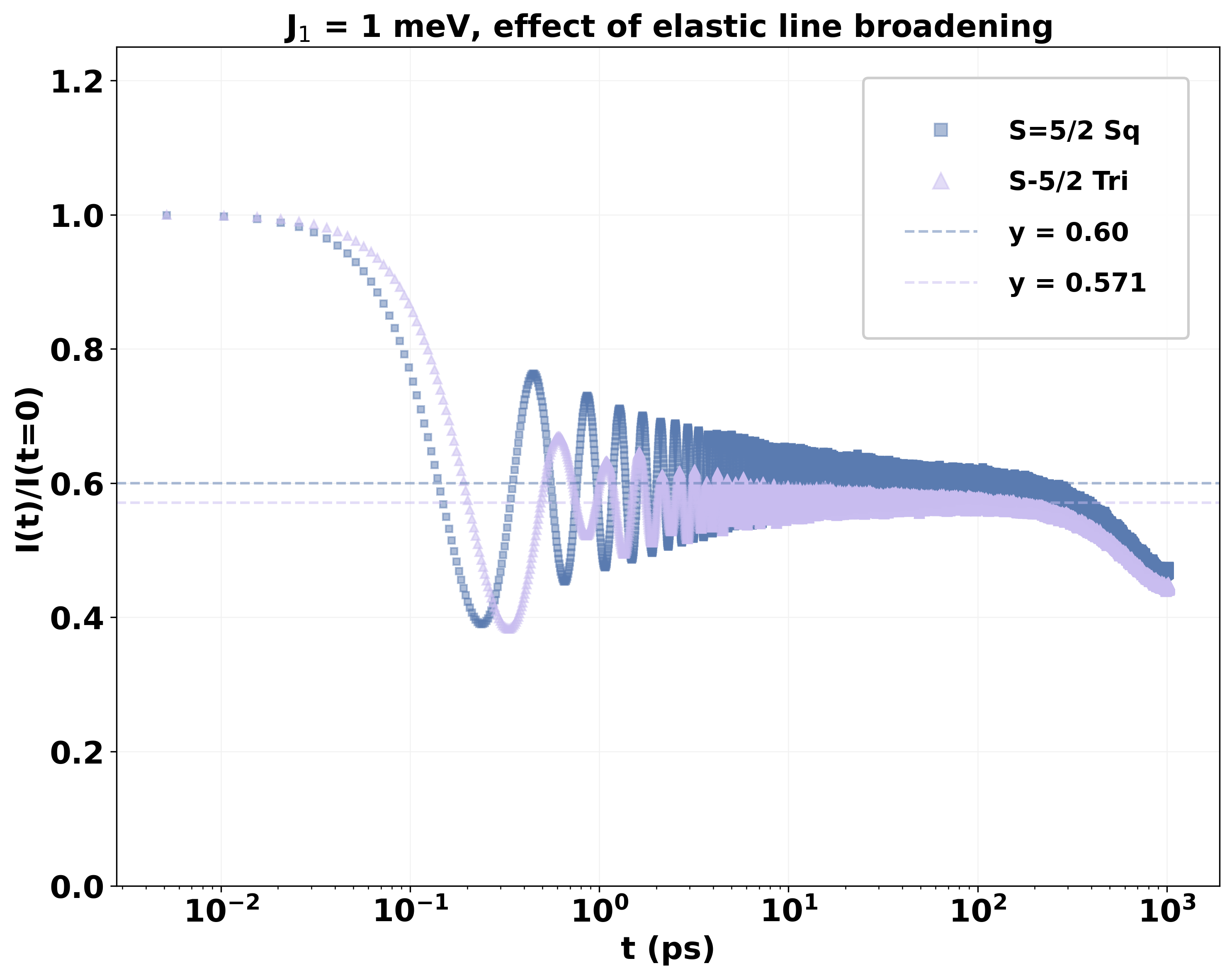}
    \caption{Effect of elastic line broadening. $\Delta E = 0.02 meV$}
    \label{fig:elasticbroadening}
\end{figure}

For example, an intrinsic broadening of $\Delta E = 0.02$~meV corresponds to a characteristic time scale $\tau \sim \hbar/\Delta E \approx 33$~ps. In this case, the nominal plateau begins to decay on time scales of order $10^{-1}$~ns, as shown in Fig.~\ref{fig:elasticbroadening}. Physically, this indicates that magnetic correlations remain effectively static on time scales shorter than $\tau$, while slower quasi-static fluctuations lead to a gradual suppression of the plateau at longer times.

More generally, the decay time scale is governed by $\hbar/\Delta E$. For high-resolution backscattering instruments with sub-$\mu$eV resolution, this decay may occur on nanosecond time scales. In contrast, for conventional spectrometers with meV-scale resolution, the decay appears only at very long times and the plateau remains well defined within the experimentally accessible time window.

%A 0.02 meV intrinsic broadening of the elastic line will cause the plateau decay from 0.1 ns. Physically speaking, it means within 33 ps, the correlations are unchanged, that is dynamics in the scale of 30 ps.
%After fast dynamic, the spin reach equilibrium and then quasi-static fluctuations slower than 30 ps.

\bibliography{aapmsamp}
\end{document}